\documentclass[conference,twoside,letterpaper,10pt]{IEEEtran}
\IEEEoverridecommandlockouts

\usepackage[pdftex]{graphicx}
\usepackage[cmex10]{amsmath}
\usepackage{cite}
\usepackage{mathtools}
\usepackage{url}

\usepackage[textfont=footnotesize,labelfont=footnotesize]{caption}
\usepackage{subcaption}
\usepackage[caption=false,labelformat=simple]{subfig}

\usepackage{color}
\usepackage{soul}
\usepackage{lipsum}

\ifCLASSINFOpdf
\else
\fi

\begin{document}
%
\title{Advanced Self-interference Cancellation and Multiantenna Techniques for Full-Duplex Radios}



%
\author{\IEEEauthorblockN{Dani Korpi\IEEEauthorrefmark{1},
Sathya Venkatasubramanian\IEEEauthorrefmark{2},
Taneli Riihonen\IEEEauthorrefmark{2}, 
Lauri Anttila\IEEEauthorrefmark{1},
Strasdosky Otewa\IEEEauthorrefmark{2},\\
Clemens Icheln\IEEEauthorrefmark{2},
Katsuyuki Haneda\IEEEauthorrefmark{2},
Sergei Tretyakov\IEEEauthorrefmark{2},
Mikko Valkama\IEEEauthorrefmark{1}, and
Risto Wichman\IEEEauthorrefmark{2}}
\\
\IEEEauthorblockA{\IEEEauthorrefmark{1}Department of Electronics and Communications Engineering, Tampere University of Technology, Finland\\ e-mail: dani.korpi@tut.fi}
\vspace{2mm}
\IEEEauthorblockA{\IEEEauthorrefmark{2}Aalto University School of Electrical Engineering, Finland\\ e-mail: sathya.venkatasubramanian@aalto.fi}
\thanks{The research work leading to these results was funded by the Academy of Finland (under the projects \#259915, \#258364 "In-band Full-Duplex MIMO Transmission: A Breakthrough to High-Speed Low-Latency Mobile Networks"), the Finnish Funding Agency for Technology and Innovation (Tekes, under the project "Full-Duplex Cognitive Radio"), the Linz Center of Mechatronics (LCM) in the framework of the Austrian COMET-K2 programme, and Emil Aaltonen Foundation.}}

\IEEEspecialpapernotice{(Invited Paper)}

\maketitle

\begin{abstract}
In an in-band full-duplex system, radios transmit and receive simultaneously in the same frequency band at the same time, providing a radical improvement in spectral efficiency over a half-duplex system. However, in order to design such a system, it is necessary to mitigate the self-interference  due to simultaneous transmission and reception, which seriously limits the maximum transmit power of the full-duplex device. Especially, large differences in power levels in the receiver front-end sets stringent requirements for the linearity of the transceiver electronics. We present an advanced architecture for a compact full-duplex multiantenna transceiver combining antenna design with analog and digital cancellation, including both linear and nonlinear signal processing.
\end{abstract}


%
\IEEEpeerreviewmaketitle

\section{Introduction}
Full-duplex communications, where a transceiver transmits and receives signals on the same frequency and time slot, has gained a lot of interest in the recent years, e.g., \cite{bliss,Choi10,Duarte12,Jain11,Bharadia13,Riihonen12,Riihonen1222}. These systems provide radical improvement in spectral efficiency over half-duplex systems, especially in relaying applications, where the frequencies can be reused. This leads to efficient utilization of the available spectrum. The main bottleneck in designing such in-band full duplex systems is the self-interference (SI) at the receiver caused by its own transmission. In order to mitigate the SI, cancellation must be done at the antenna, radio frequency (RF), and digital domains.

Although previous work has been done to cancel the SI, the achieved cancellation levels need to be improved for practical implementations and robust solutions are necessary. In this regard, we propose different techniques to mitigate the SI signal in the antenna and digital domains. Since it is expected that relays are the first candidates for such full-duplex systems, in Section~\ref{sec:antenna} we suggest isolation improvement techniques in the antenna domain for a compact relay using loops for field suppression. Section~\ref{sec:digital} discusses the nonlinearities occurring in the receiver chain of a full-duplex transceiver and proposes an algorithm to estimate and cancel a nonlinearly distorted SI signal. Finally, the conclusions are drawn in Section~\ref{sec:conc}.

\section{Antenna Isolation Improvement for a Compact Relay}
\label{sec:antenna}

\begin{figure}[!b]
\centering
\includegraphics[width=0.8\columnwidth]{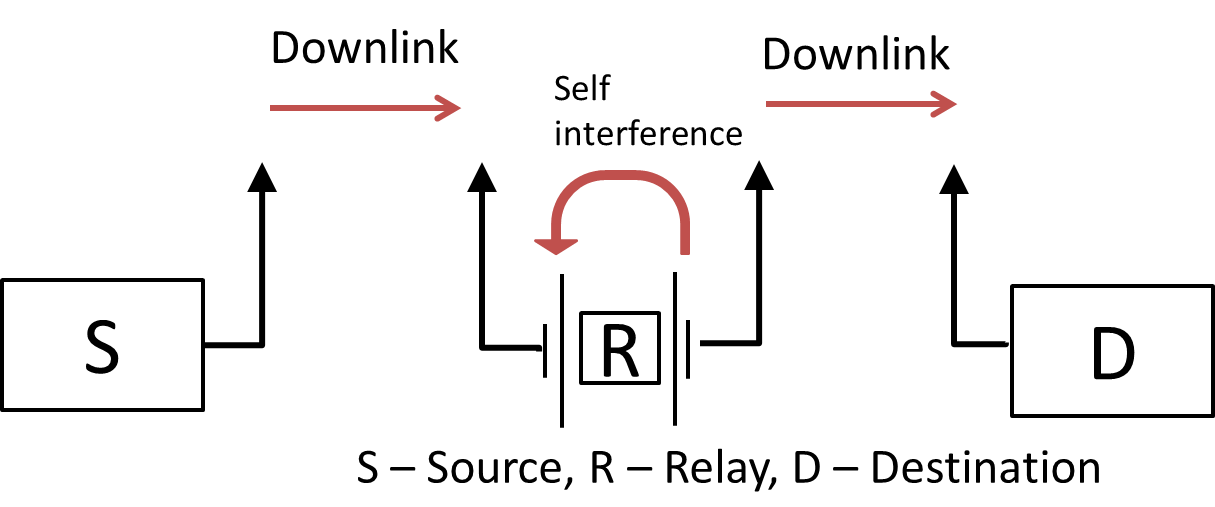}
\caption{In-band full-duplex relay.}
\label{fig:relay_diagram}
\end{figure}

In this section, we consider the scenario of a compact relay with back-to-back antennas operating at the same frequency and time slot, as depicted in Fig.~\ref{fig:relay_diagram}. The full-duplex relay operates at the same frequency for the the source-to-relay, and relay-to-destination downlinks \cite{Riihonen124}. The uplink and downlink frequencies or timeslots are different. These compact relays can be used in outdoor-to-indoor scenarios or installed on lamp-posts as hotspots to improve the dead-spot coverage and provide higher capacity to the end user. 

\begin{figure}[!t]
\centering
\includegraphics[width=0.75\columnwidth, trim = 0mm 0mm 25mm 0mm]{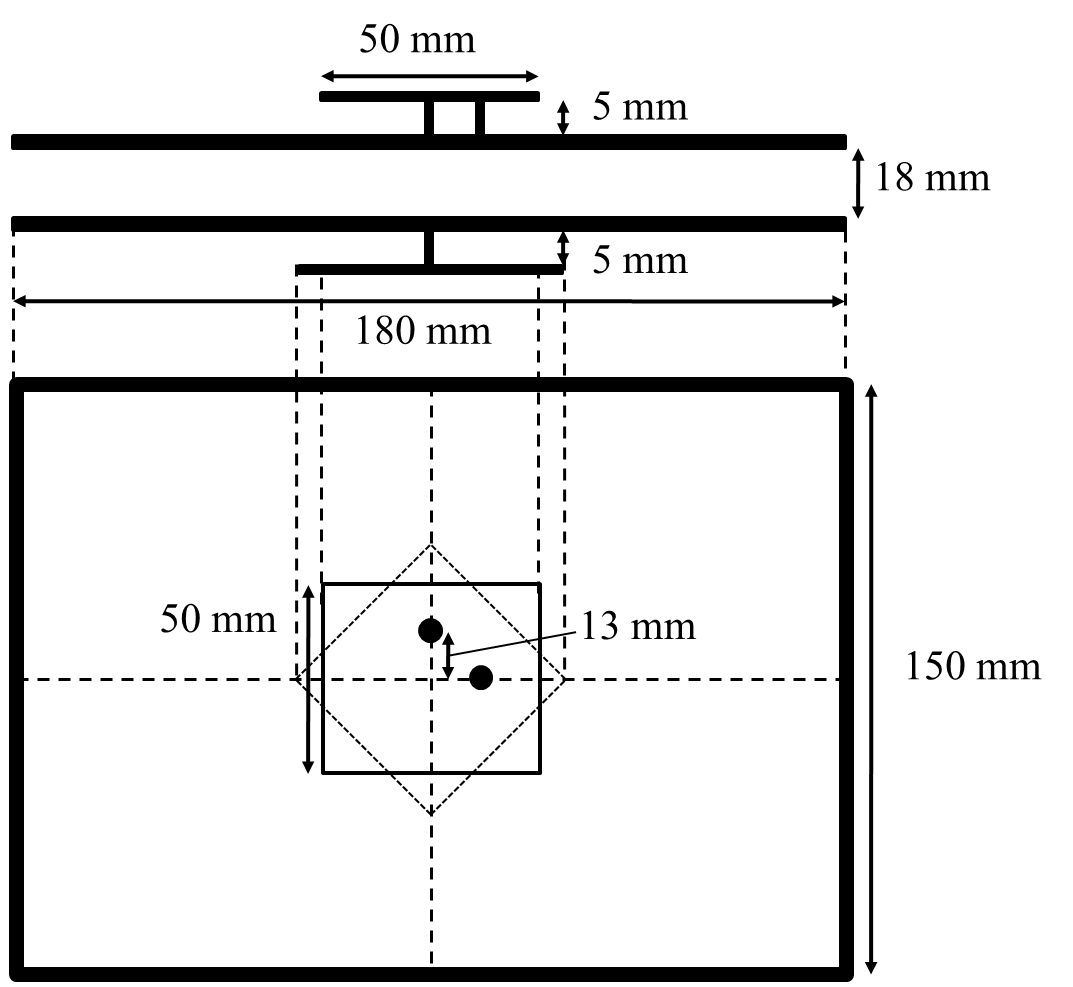}
\caption{Design of the compact relay.}
\label{fig:relay_reference_antenna}
\vspace{-4mm}
\end{figure}

Based on the earlier work in \cite{RefWorks:13} on measurements of the loop-back interference channel, the following specifications were considered for designing the compact relay:
\begin{itemize}
\item the relay dimension is comparable to that of a wireless access point, and
\item the relay antennas operate at 2.6 GHz with a minimum bandwidth of 100 MHz.
\end{itemize}
We consider two dual-polarized patch antennas placed back-to-back as shown in Fig. \ref{fig:relay_reference_antenna}. The antenna on one ground plane is used for the source-to-relay link and the antenna on the second ground plane is used for the relay-to-destination link. The antennas on one side of the ground plane are tilted by $45^{\circ}$ to improve the isolation. As the separation distance between the transmit (TX) and receive (RX) antennas is electrically small, i.e., $0.156\lambda$ at 2.6 GHz, we need to analyze the near fields created by the antennas to calculate the isolation between them. The isolation level obtained is the SI cancellation achieved between the TX and RX antennas.

In order to improve the isolation for this antenna, different methods can be used, including the use of wavetraps \cite{wavetrap}, band-gap structures \cite{ebg}, slots on the ground plane \cite{slots} and RF absorbers. In this paper, we investigate the use of loops to cancel the magnetic field and hence improve the isolation between the TX and RX antennas. When investigating the electromagnetic fields around the antenna using HFSS simulations, which uses the finite element method \cite{hfss}, it was observed that the near-field  impedance, which is the ratio of the transverse electric and magnetic fields, was less than $120\pi \text{ } \Omega$. This indicates that the magnetic fields are dominant in the region between the ground planes. Hence, cancellation of the magnetic fields can provide an improvement in the isolation.

\begin{figure}[!t]
\centering
\includegraphics[width=0.75\columnwidth, trim = 0mm 0mm 20mm 0mm]{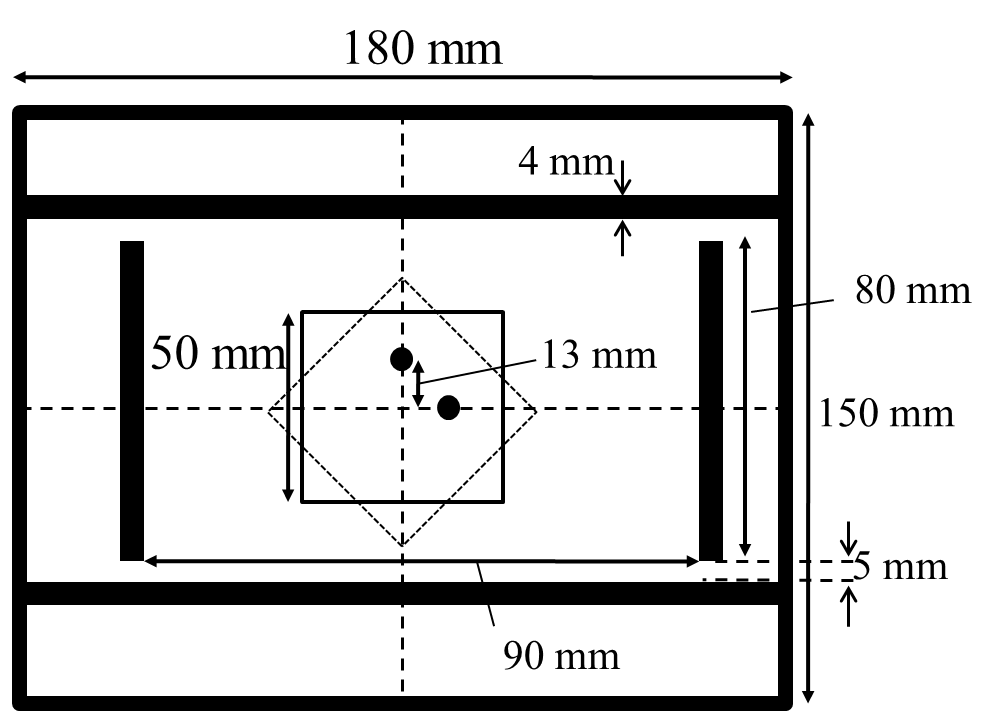}
\caption{Top view of antenna structure with loops for field cancellation.}
\label{fig:relay_with_loop}
\end{figure}

\begin{figure}[!t]
\centering
\begin{subfigure}[b]{\columnwidth}
\includegraphics[width=0.96\columnwidth]{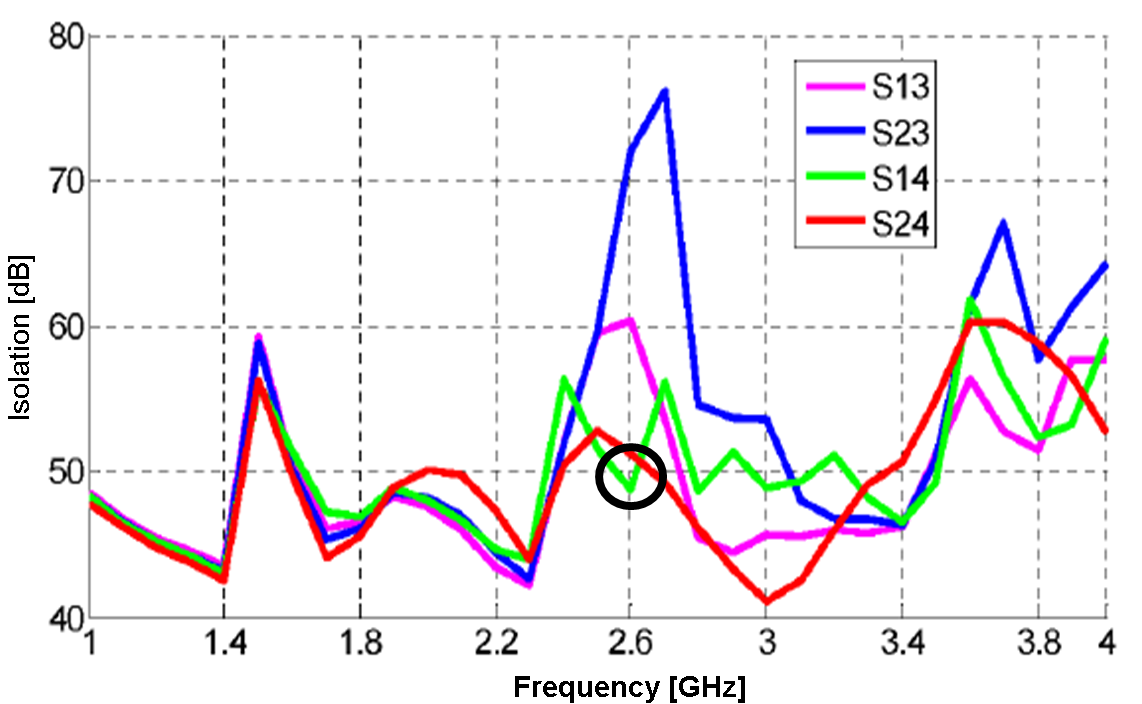}
\caption{Isolation without loops.}
\label{fig:isolation_ref}
\end{subfigure}

\begin{subfigure}[b]{\columnwidth}
\centering
\includegraphics[width=\columnwidth]{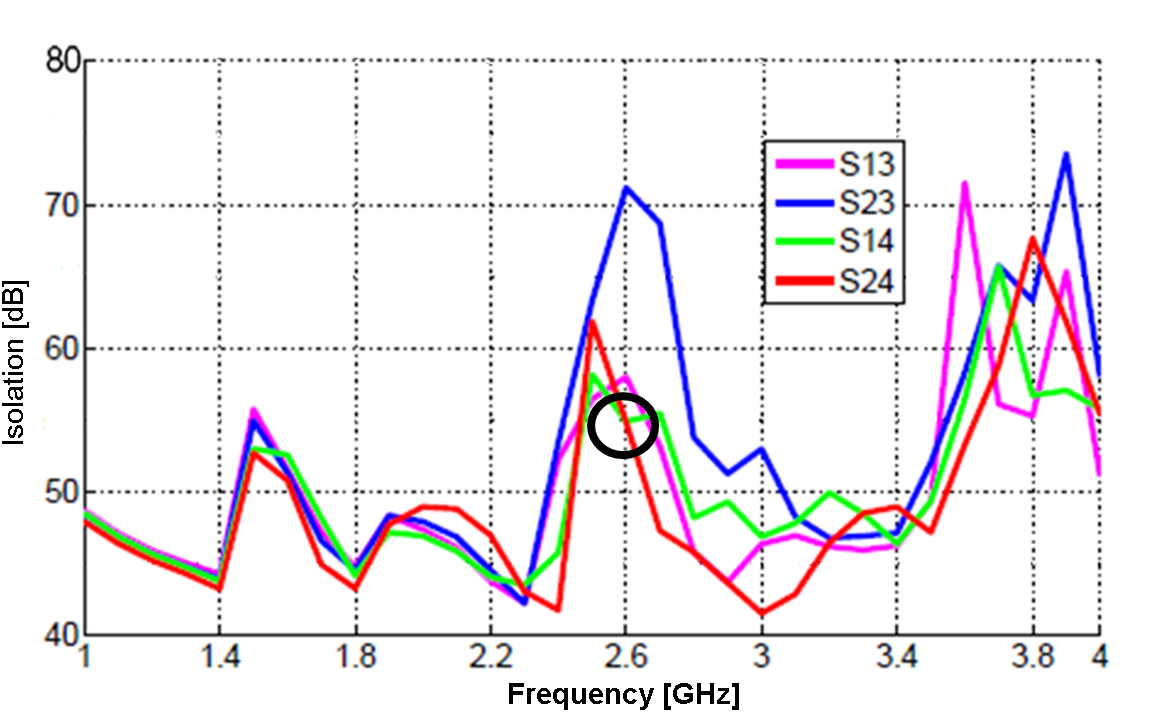}
\caption{Isolation with loops for field cancellation.}
\label{fig:isolation_with_loops}
\end{subfigure}
\caption{The isolations between the antenna ports with and without loops.}
\end{figure}

To this end, loops were designed on the ground plane on one side of the relay as shown in Fig. \ref{fig:relay_with_loop}. The concept of using loops stems from Faraday's law of electromagnetic induction and Lenz's law \cite{cheng}. The induced currents in the loops create magnetic fields, which oppose the change in the flux producing it, thereby partially cancelling the local reactive magnetic fields produced by the antenna. Figs. \ref{fig:isolation_ref} and \ref{fig:isolation_with_loops} show the isolation of the relay antenna without the field cancellation loops and with the loops on the ground plane, respectively. It can be observed that, at the operating frequency of 2.6 GHz, we have an improvement of 6 dB in the worst case isolation shown by the black circles, indicating a minimum isolation of 55 dB between the transmit and receive antennas. The isolation can be further improved  by combining different techniques mentioned above for practical deployment of full-duplex relays. 

\section{Cancellation of RX-induced Nonlinear Distortion}
\label{sec:digital}
In addition to antenna attenuation and possible analog SI cancellation, usually additional SI attenuation is required in the digital domain. This is due to the fact that the power of the SI is usually not attenuated sufficiently in the analog domain and thus, without further SI suppression in the digital domain, the decrease in the receiver chain signal-to-interference-plus-noise ratio (SINR) would be intolerable. Traditionally, the SI suppression in the digital domain, or digital cancellation, has been performed using linear processing methods. This means that the SI coupling channel, including the effects of transmit chain, actual coupling channel between the antennas, and receiver chain, is estimated with linear channel estimation methods \cite{Choi10,Jain11,Duarte12,Sahai11}. However, the power of the own transmit signal entering the receiver chain may be in the order of 60--80 dB higher than the power of the weak received signal of interest. Thus, unless assuming highly linear receiver components, the power the total signal entering the receiver chain may be high enough to drive the components out of their linear operating region. This means that the total signal will be distorted nonlinearly in the receiver chain, and it is not possible to regenerate and suppress the SI signal using a linear SI channel estimate.

There have recently been studies on nonlinear SI cancellation, but none of them concentrate on RX-induced nonlinear distortion \cite{Anttila13,Ahmed13,Bharadia13}. In this paper, it will be shown that nonlinear distortion produced in the receiver chain can indeed limit the performance of a full-duplex transceiver, and this provides strong motivation to develop an algorithm capable of attenuating also a nonlinearly distorted SI signal. Thus, we describe an algorithm capable of estimating and modeling nonlinear distortion, which will increase the amount of achieved digital cancellation in a typical full-duplex transceiver.

\subsection{MIMO Full-duplex Transceiver Model}

\begin{figure*}[!t]
\centering
\includegraphics[width=\textwidth]{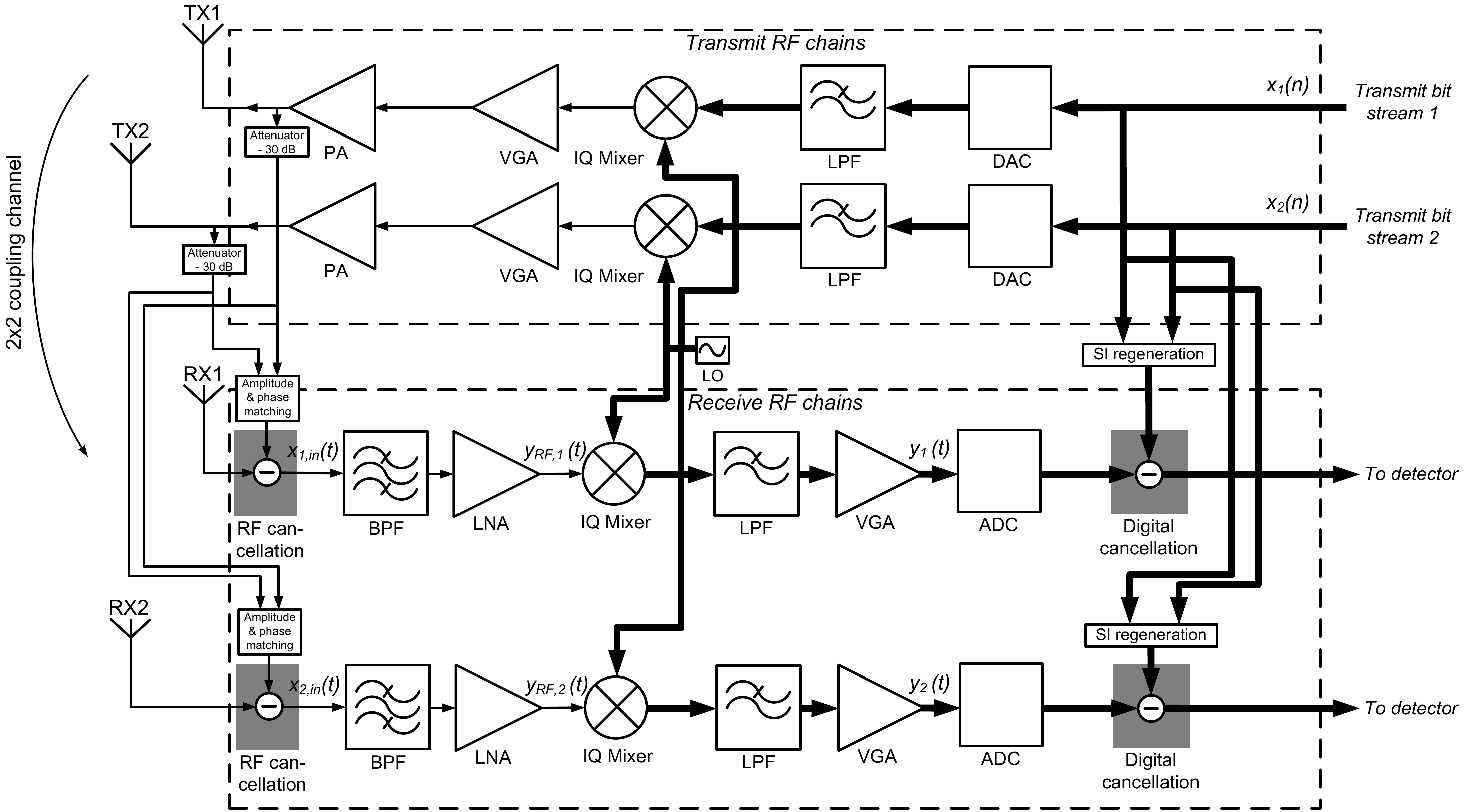}
\caption{A block diagram of the assumed 2x2 MIMO full-duplex transceiver.}
\label{fig:block_diagram}
\end{figure*}

To allow for a more general scenario, in this paper a MIMO direct-conversion full-duplex transceiver is assumed. Thus, there are several transmitters and receivers operating simultaneously in the considered transceiver. This means that the total SI signal coupling to a single receiver antenna consists of the sum of all the transmitted signals. An ideal baseband model of this type of a MIMO full-duplex transceiver is presented in, e.g., \cite{Riihonen12}. For the different calculations and modeling, a 2x2 MIMO is considered in this paper, meaning that the full-duplex transceiver has two transmit antennas, two receiver antennas, and four RF chains. A block diagram of this transceiver is shown in Fig.~\ref{fig:block_diagram}.

The considered MIMO full-duplex transceiver model has two active SI cancellation stages, in addition to passive antenna attenuation. At the input of each receiver chain, RF cancellation is performed, where the transmitted signals are subtracted from the received signal. After the actual receiver chain, further SI cancellation is performed in the digital domain. This is done by estimating the channel of the SI signal and then regenerating it based on the channel estimate. The novel nonlinear SI cancellation algorithm proposed in this paper is in essence increasing the precision of the regenerated SI signal, and thus increasing the amount of achievable digital SI cancellation when operating with practical nonlinear RF components.

To model a realistic scenario, it is assumed that the receiver chains of the considered MIMO full-duplex transceiver have typical low to medium-cost components. This means that the linearity of the amplifiers and IQ mixer is not as high as for the more expensive state of the art components. The considered parameter values are listed in Tables~\ref{table:system_parameters} and~\ref{table:parameters}. Since the objective of this analysis is to determine the effect of the receiver chain nonlinearities, it is assumed that the PAs of the transmit chains are completely linear. This allows for a detailed analysis of only the RX-induced nonlinear distortion.

\begin{table}[!t]
\renewcommand{\arraystretch}{1.3}
\caption{System level and general parameters of the 2x2 MIMO full-duplex transceiver.}
\label{table:system_parameters}
\centering
\begin{tabular}{|c||c|}
\hline
\textbf{Parameter} & Value\\
\hline
SNR requirement & 10 dB \\
\hline
Bandwidth & 12.5 MHz\\
\hline
Receiver noise figure & 4.1 dB\\
\hline
Sensitivity & -88.9 dBm\\
\hline
Received signal power & -83.9 dBm\\
\hline
Antenna separation & 40 dB\\
\hline
RF cancellation & 20 dB\\
\hline
ADC bits & 12\\
\hline
PAPR & 10 dB\\
\hline
PA gain & 27 dB\\
\hline
\end{tabular}
\end{table}

\begin{table}[!t]
\renewcommand{\arraystretch}{1.3}
\caption{Parameters for the relevant components of the receiver chains.}
\label{table:parameters}
\centering
\begin{tabular}{|c||c||c||c||c|}
\hline
\textbf{Component} & \textbf{Gain (dB)} & \textbf{IIP2 (dBm)} & \textbf{IIP3 (dBm)} & \textbf{NF (dB)}\\
\hline
LNA (RX) & 25& 43 & -15 & 4.1\\
\hline
Mixer (RX)& 6 & 42 & 15 & 4\\
\hline
VGA (RX) & 0-69 & 43 & 10 & 4\\
\hline\end{tabular}
\end{table}

\subsection{RX-induced Nonlinearities in a MIMO Full-Duplex Transceiver}

To obtain some insight into the effect of receiver chain induced nonlinear distortion occuring in a full-duplex transceiver, simplified system calculations are performed. This reveals the approximate power levels of the different signal components after linear digital cancellation for one receiver chain. In this analysis, the power levels are calculated with respect to transmit power of a single transmit antenna under the assumption that all the transmit chains use the same transmit power. A detailed derivation of all the necessary equations for a SISO scenario can be found in \cite{Korpi13}. In this analysis, it is assumed that digital cancellation attenuates the linear SI component slightly below the thermal noise floor. This is a reasonable assumption in this context, as the purpose of this simplified example is to illustrate the relative power level of the receiver chain induced nonlinear distortion.

\begin{figure}[!t]
\centering
\includegraphics[width=\columnwidth]{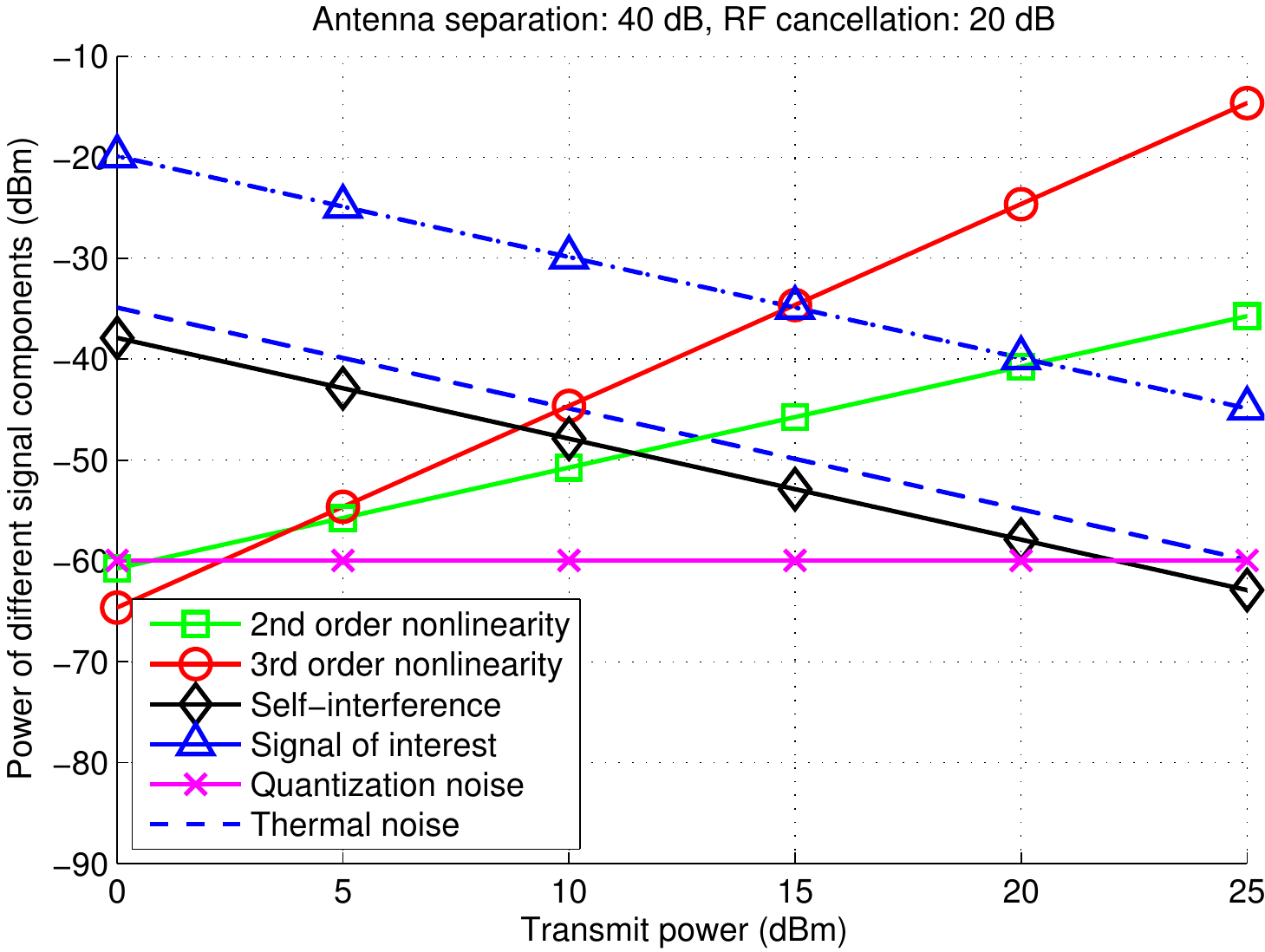}
\caption{An example plot of the power levels of the different signal components at receiver chain detector input with respect to transmit power.}
\label{fig:p_all_example}
\end{figure}

The different power levels of the signal components for one receiver chain, corresponding to the chosen parameters, are shown in Fig.~\ref{fig:p_all_example}. It can be observed that the 2nd- and 3rd-order RX-chain induced nonlinearities are the most significant distortion components with transmit powers above 10 dBm. Furthermore, it is evident that the strength of the nonlinear distortion is sufficient to significantly degrade the SINR with high transmit powers. Thus, the capability to model and attenuate also a nonlinearly distorted SI signal will provide performance gain for a full-duplex transceiver, especially with higher transmit powers.

\subsection{Nonlinear SI Cancellation Algorithm and Parameter Estimation}

The nonlinear distortion occuring in the receiver chain of a full-duplex transceiver can be approximated with polynomials. There have been previous studies on modeling the nonidealities of receiver chains \cite{Grimm13}, and thus the models for the nonlinearities are not derived in this paper. Instead, a model presented in \cite{Grimm13} is used, with the addition of a 2nd-order nonlinear term, produced by the baseband components and the IQ mixer. The baseband equivalent model for the low-noise amplifier is now $y_{RF,i}(t) = k_{LNA}x_{i,in}(t)+\alpha_{i} |x_{i,in}(t)|^2 x_{i,in}(t)$, where $x_{i,in}(t)$ is the input signal of the $i$th receiver chain. The combined baseband equivalent model for the IQ mixer and variable gain amplifier can be expressed as $y_{i}(t) = k_{BB}y_{RF,i}(t)+\beta_{i} |y_{RF,i}(t)|^2 + \gamma_{i} \left[y^{\ast}_{RF,i}(t)\right]^3$, where $()^{\ast}$ denotes the complex conjugate. During a training period when there is no actual signal of interest present, the overall signal model at the digital baseband of one receiver chain is as follows:
\begin{align}
	&y_i(n) = a_{i,1} x_{i,in}(n) + a_{i,2} |x_{i,in}(n)|^2 \nonumber\\
	&+ a_{i,3} |x_{i,in}(n)|^2 x_{i,in}(n)+ a_{i,4} \left[x_{i,in}^{\ast}(n)\right]^3 + z_i(n) \text{,} \label{eq:sig_model}
\end{align}
where $x_{i,in}(n)$ is now the total SI signal at the input of the $i$th receiver chain, $a_{i,k}$ are the combined coefficients of the different linear and nonlinear terms, and $z_i(n)$ represents the other noise sources. Here, only terms up to 3rd order are considered, as it was observed that the higher-order terms are insignificant with the chosen parameters. Based on \eqref{eq:sig_model}, it is possible to calculate the coefficients of the different terms by stacking the sampled signals to vectors of observation period~$N$.

However, for that, the input signal of the receiver chain must first be determined. This can be done by obtaining an estimate of the linear SI coupling channel with a low transmit power. In that case, the receiver chain can be assumed to be completely linear, and linear channel estimation provides an accurate result. The signal model during a training period, assuming a low transmit power, can be written as follows:
\begin{align}
	y_i(n) = \sum_{j=1}^{N_{tx}} h_{ij}(n) \star x_j(n) + z_i(n) \text{,} \label{eq:lowtx_model}
\end{align}
where $N_{tx}$ is the number of transmit antennas and $h_{ij}(n)$ is the effective channel experienced by the $j$th transmit signal $x_{j}(n)$ when propagating to the $i$th receiver, including the effects of linear RF cancellation and the receiver chain. Because of the MIMO context, the total SI signal is a sum of all the transmit signals, each of which experiences a slightly different coupling channel. With vector notations, and using again an observation period of $N$ samples, \eqref{eq:lowtx_model} can be rewritten as follows:
\begin{align}
	\mathbf{y}_i &= \sum_{j=1}^{N_{tx}} \mathbf{X}_j \mathbf{h}_{ij} + \mathbf{z}_i \nonumber\\
	&= \begin{bmatrix}\mathbf{X}_1 & \mathbf{X}_2 & \cdots &  \mathbf{X}_{N_{tx}}\end{bmatrix} \begin{bmatrix}
	\mathbf{h}_{i1}\\
	\mathbf{h}_{i2}\\
	\vdots \\
	\mathbf{h}_{iN_{tx}} 
	\end{bmatrix} + \mathbf{z}_i \nonumber\\
	&= \mathbf{X}_{tot} \mathbf{h}_{i,tot} + \mathbf{z}_i \text{,} \label{eq:y_mtrx}
\end{align}
where
\begin{align*}
	&\mathbf{h}_{ij} = \begin{bmatrix} h_{ij}(0) & h_{ij}(1) & \cdots &  h_{ij}(M-1) \end{bmatrix}^{T} \text{,}\\
	&\mathbf{h}_{i,tot} = \begin{bmatrix}\mathbf{h}^T_{i1} & \mathbf{h}^T_{i2} & \cdots & \mathbf{h}^T_{iN_{tx}}\end{bmatrix}^{T} \text{,}\\
	&\mathbf{X}_{tot} = \begin{bmatrix}\mathbf{X}_1 & \mathbf{X}_2 & \cdots & \mathbf{X}_{N_{tx}}\end{bmatrix} \text{,}
\end{align*}
and $\mathbf{X}_j$ is a covariance windowed convolution matrix of the form
\begin{align*}
\mathbf{X}_j =  \begin{bmatrix}
  x_j(M-1) & x_j(M-2) & \cdots & x_j(0) \\
  x_j(M) & x_j(M-1) & \cdots & x_j(1) \\
  \vdots  & \vdots  & \ddots & \vdots  \\
  x_j(N-1)& x_j(N-2) & \cdots & x_j(N-M)
 \end{bmatrix} \text{.}
\end{align*}
Here, $M$ denotes the length of the channel estimate. This same signal model is also presented in \cite{Riihonen12} with matrix notations. Now, the SI channel responses for each receiver can be calculated by, e.g., linear least squares as follows:
\begin{align}
	\mathbf{\hat{h}}_{i,tot} = (\mathbf{X}^H_{tot} \mathbf{X}_{tot})^{-1}\mathbf{X}^H_{tot} \mathbf{y}_i \text{,} \label{eq:linear_ls}
\end{align}
where $\mathbf{y}_i = \begin{bmatrix} y_i(M-1) & y_i(M) & \cdots &  y_i(N-1) \end{bmatrix}^{T}$, and $()^{H}$ denotes the Hermitian transpose. Since low transmit power is used during this training period, the effect of the receiver chain can be approximated by merely a scalar multiplication. Thus, $\mathbf{\hat{h}}_{i,tot}$ includes in fact estimates of the channels experienced by the transmit signals before the $i$th receiver chain, scaled by the amplification of the receiver chain in question. Assuming that the channels change sufficiently slowly, these channel estimates can be used to form an estimate of the receiver chain input signal during another training period, where the coefficients of the nonlinear terms are then calculated.

To determine the coefficients of the nonlinear terms, the signal model presented in \eqref{eq:sig_model} must first be written with vector notations as follows:
\begin{align}
	\mathbf{y}_i = \mathbf{X}_{nl,i} \mathbf{a}_i +\mathbf{z}_i  \text{,} \label{eq:y_nlmtrx}
\end{align}
where
\begin{align}
&\mathbf{X}_{nl,i} = \begin{bmatrix}
  \mathbf{x}_{i,in} & |\mathbf{x}_{i,in}|^2 & |\mathbf{x}_{i,in}|^2 \mathbf{x}_{i,in} & \left[\mathbf{x}_{i,in}^{\ast}\right]^3
 \end{bmatrix} \text{,} \nonumber\\
& \mathbf{a}_i = \begin{bmatrix} a_{i,1} & a_{i,2} & a_{i,3} &  a_{i,4} \end{bmatrix}^{T} \text{,} \nonumber\\
& \mathbf{x}_{i,in} = \begin{bmatrix} x_{i,in}(M-1) & x_{i,in}(M) & \cdots &  x_{i,in}(N-1) \end{bmatrix}^{T} \text{,} \nonumber
\end{align}
and all the transformations in the matrix $\mathbf{X}_{nl,i}$ are performed element-wise to the column vector $\mathbf{x}_{i,in}$. As this signal model is linear in parameters, the coefficients $\mathbf{a}_i$ can be calculated with linear least squares as follows.
\begin{align}
	\mathbf{\hat{a}}_i = (\mathbf{X}^H_{nl,i}\mathbf{X}_{nl,i})^{-1}\mathbf{X}^H_{nl,i} \mathbf{y}_i \text{.} \label{eq:nonlinear_ls}
\end{align}
Since this estimation procedure is performed during a training period with an arbitrary transmit power, the signal $\mathbf{y}_i$ might also be nonlinearly distorted, unlike previously.

Now, using the linear channel estimates $\mathbf{\hat{h}}_{i,tot}$ and the known transmit signals $x_j(n)$ to form an estimate of the receiver chain input signal $\mathbf{x}_{i,in}$, the estimates for the coefficients of the nonlinear terms $\mathbf{a}_i$ can be calculated with~\eqref{eq:nonlinear_ls}. With these estimates, it is possible to reconstruct the SI signal when using the same transmit power that was used during the estimation of the coefficients. Thus, even a nonlinearly distorted SI signal can be suppressed in the digital domain by knowing only the transmitted symbols, as long as the overall channel remains approximately constant.

\subsection{Waveform Simulations}

\begin{table}[!t]
\renewcommand{\arraystretch}{1.3}
\caption{Additional parameters for waveform simulations.}
\label{table:ofdm_param}
\centering
\begin{tabular}{|c||c|}
\hline
\textbf{Parameter} & \textbf{Value}\\
\hline
Constellation & 16-QAM\\
\hline
Number of subcarriers & 64\\
\hline
Number of data subcarriers & 48\\
\hline
Guard interval & 25 \% of symbol length\\
\hline
Sample length & 15.625 ns\\
\hline
Symbol length & 4 $\mu$s\\ 
\hline
Oversampling factor & 4\\
\hline
\end{tabular}
\vspace{-4mm}
\end{table}

To evaluate the performance of the proposed algorithm, waveform simulations are performed. The same transceiver model is used as presented in Fig.~\ref{fig:block_diagram}, with the parameters shown in Tables~\ref{table:system_parameters} and~\ref{table:parameters}. Furthermore, in the simulations an OFDM signal is assumed, with the parameters presented in Table~\ref{table:ofdm_param}. Basically, these parameters correspond to a WLAN system. In addition, the two transmit data streams are independent. The lengths of the channel estimate ($M$) and observation period ($N$) are chosen to be $5$ and $10000$, respectively. In the simulations, the transmit power is varied with 2.5 dB intervals, and 20 realizations are calculated for each transmit power. The SINR for each transmit power is calculated as the average value of these realizations.

\begin{figure}[!t]
\centering
\includegraphics[width=\columnwidth]{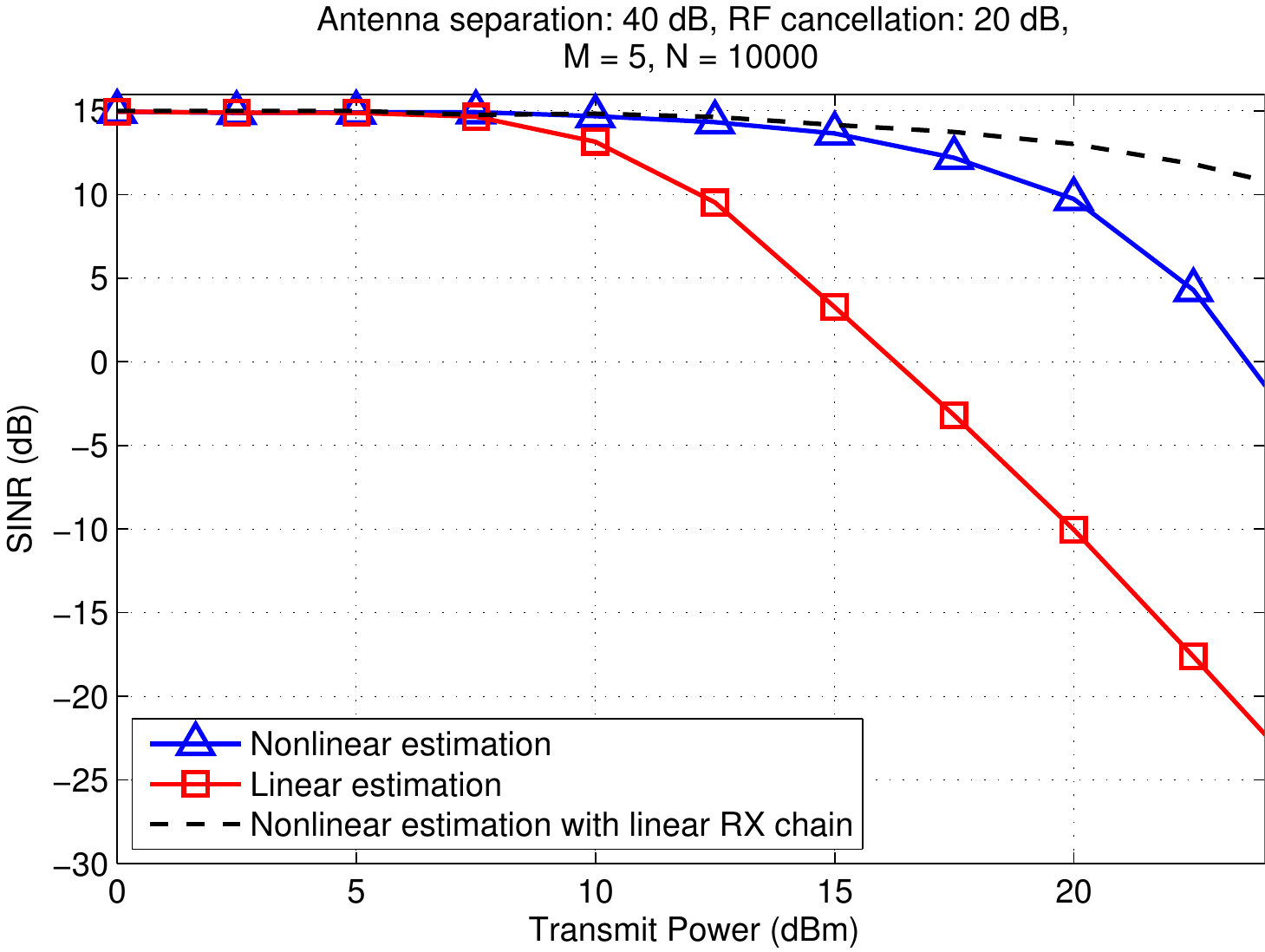}
\caption{The simulated SINRs corresponding to one receiver chain.}
\label{fig:sinr_simul}
\vspace{-5mm}
\end{figure}

The simulated SINRs corresponding to one receiver chain, with respect to the transmit power of a single antenna, are presented in Fig.~\ref{fig:sinr_simul}. The SINR is shown for three scenarios: using only linear digital SI cancellation for a nonlinear receiver chain, and using the proposed nonlinear digital SI cancellation algorithm for nonlinear and linear receiver chains. There are two significant observations that can be made from the figure. Firstly, it can be observed that the proposed nonlinear SI cancellation algorithm increases the achievable SINR significantly with higher transmit powers, when compared to using only linear digital cancellation. Secondly, even with a completely linear receiver chain, the proposed algorithm is not able to achieve the ideal SINR of 15 dB, which corresponds to a situation where the SI signal is cancelled completely. The reason for this lies in the estimation procedure of the linear SI channel, which is done with a lower transmit power to ensure the linearity of the receiver chain. The lower transmit power means that the power of the SI signal is also lower with respect to the noise floor, and this results in a more noisy SI channel estimate than could be achieved with a higher SI power. Thus, the estimate of the receiver chain input signal is also noisy, which decreases the achievable SINR with higher transmit powers, where very precise channel estimates are required to suppress the SI signal below the noise floor. Anyway, the proposed nonlinear digital SI cancellation algorithm still achieved higher SINRs than when using linear processing methods, indicating that it is indeed beneficial to model also the nonlinearities occuring in the receiver chain.

\vspace{-2.5mm}
\section{Conclusion}
\label{sec:conc}
\vspace{-1mm}
We have presented methods to mitigate the self-interference in the antenna and digital domains. Antenna isolation level of 55 dB was achieved in a compact relay using loops for field suppression. In the digital domain, an algorithm capable of cancelling a nonlinearly distorted self-interference signal was proposed, assuming a nonlinear receiver chain. Although these techniques provide solutions to attenuate self-interference, further work is necessary to design cancellation techniques for practical implementations of such systems.






%
\ifCLASSOPTIONcaptionsoff
  \newpage
\fi

\bibliographystyle{IEEEtran}
\bibliography{./IEEEabrv,./IEEEref}

\end{document}